\documentclass[preprint,12pt]{elsarticle}

\usepackage{amssymb}

\newcommand{\dphi}{\Delta\phi}
\newcommand{\pt}{p_{T}}
\journal{Nuclear Physics A}

\begin{document}

\begin{frontmatter}



\title{d+Au Hadron Correlation Measurements at PHENIX}


\author{Anne M. Sickles, for the PHENIX Collaboration}
\ead{anne\@bnl.gov}
\address{Physics Department, Brookhaven National Laboratory, Upton NY}

\begin{abstract}
In these proceedings, we discuss recent results from d+Au collisions in 
PHENIX  ridge related measurements and their possible
hydrodynamic origin.
We present the  $v_2$ at midrapidity and
measurements of the pseudorapidity dependence of the
ridge, distinguishing between the d-going and
Au-going directions.
We investigate the possible geometrical origin
by comparing $v_2$ in d+Au to that in p+Pb, Au+Au and Pb+Pb collisions.
Future plans to clarify the
role of geometry in small collision systems at RHIC are discussed.

\end{abstract}


\end{frontmatter}

In heavy ion collisions at both RHIC and the LHC the properties of the created matter
are understood to be described by hydrodynamics with a very small shear viscosity
to entropy density ratio, $\eta/s$~\cite{Adare:2011tg}.  
The value of $\eta/s$ is constrained via measurements of Fourier coefficients of the azimuthal
distribution of particles ($v_N$ where $N$ is the order of the Fourier coefficient).
Measurements of the particle pair correlations in heavy ion collisions have been well
described by products of the same $v_N$ at the appropriate $p_T$~\cite{Aamodt:2011by,ATLAS:2012at}.  One prominent feature of these correlation
functions is the so-called {\it ridge}, a long range in pseudorapidity, small $\dphi$
correlation resulting from the sum of positive $v_N$ from hydrodynamic flow~\cite{Abelev:2009af,Aamodt:2011by}. 

Surprisingly, a similar long range correlation was observed in very high multiplicity p+p
collisions at the LHC~\cite{Khachatryan:2010gv} where a hydrodynamical system was not
generally expected to be created.  
Recently, at the LHC a double ridge structure,
with long range correlations at $\Delta\phi=$~0 and $\pi$ has been observed also in p+Pb collisions
at 5.02~TeV~\cite{Abelev:2012ola,Aad:2012gla}.  
This feature can be largely described by a $\cos2\Delta\phi$ modulation.
Extractions of $v_2$ yielded values with a similar magnitude to those
in heavy ion collisions~\cite{Abelev:2012ola,Aad:2012gla}.
Obviously, such similar behavior between A+A and p(d)+A collisions is suggestive of
a similar physical origin.  This was quite surprising considering the small
size of the overlap region between the two nuclei  in p(d)+A collisions.
The observation of a double ridge structure, of course, does not prove hydrodynamic behavior;
in fact, similar effects were expected within the Color Glass Condensate model~\cite{Dusling:2012wy}.
Therefore, experimentally, it is of great interest to investigate the nature of the
correlations observed in p(d)+A and to compare measurements in p(d)+A to what is
known from A+A collisions.
Approximately 1.6 billion d+Au collisions
at $\sqrt{s_{NN}}$~=~200~GeV was taken in 2008.  Here we report on measurements by the PHENIX collaboration
using that data.

\section{Midrapidty Correlations \& $v_2$}

\begin{figure}
\includegraphics[width=\textwidth]{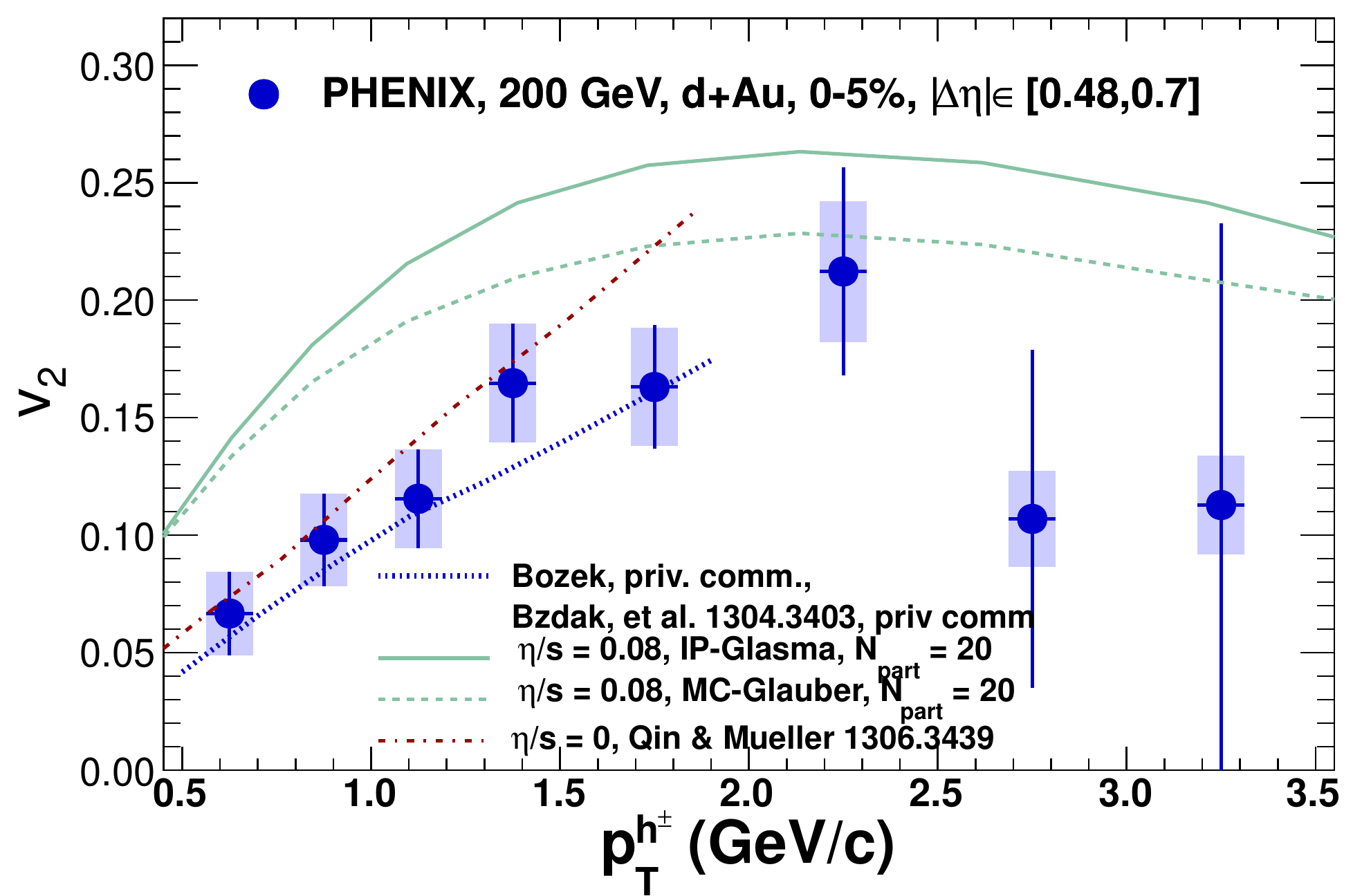}
\caption{$v_2$ as a function of $p_{T}$ for midrapidity hadrons
in the 5\% most central d+Au collisions.  Also shown on the plot are
hydrodynamic calculations from Refs.~\cite{Bozek:2011if,Bozek:privatecomm,Bzdak:2013zma,Qin:2013bha}.
Figure is from Ref.~\cite{Adare:2013piz}.}
\label{fig:v2}
\end{figure}

PHENIX has measured the $v_2$ of charged hadrons at midrapidity in central d+Au collisions using
two-particle correlations.
Correlations from jets are removed under the assumption that the conditional yields are unmodified from
peripheral d+Au events.
Resulting $v_2$ values as a function of $p_T$ are shown in Fig.~\ref{fig:v2}~\cite{Adare:2013piz}. 
The $v_2$ rises with $\pt$  reaching a maximal value of about 15\%.  
Also shown on the plot are hydrodynamic calculations from three 
groups~\cite{Bozek:2011if,Bozek:privatecomm,Bzdak:2013zma,Qin:2013bha}.  All three
calculations agree well with the data.
Refs.~\cite{Bozek:2011if,Bozek:privatecomm,Bzdak:2013zma} use $\eta/s$~=~0.08,
while the calculation in Ref.~\cite{Qin:2013bha} is for ideal hydrodynamics.

\begin{figure}
\centering
\includegraphics[width=0.7\textwidth]{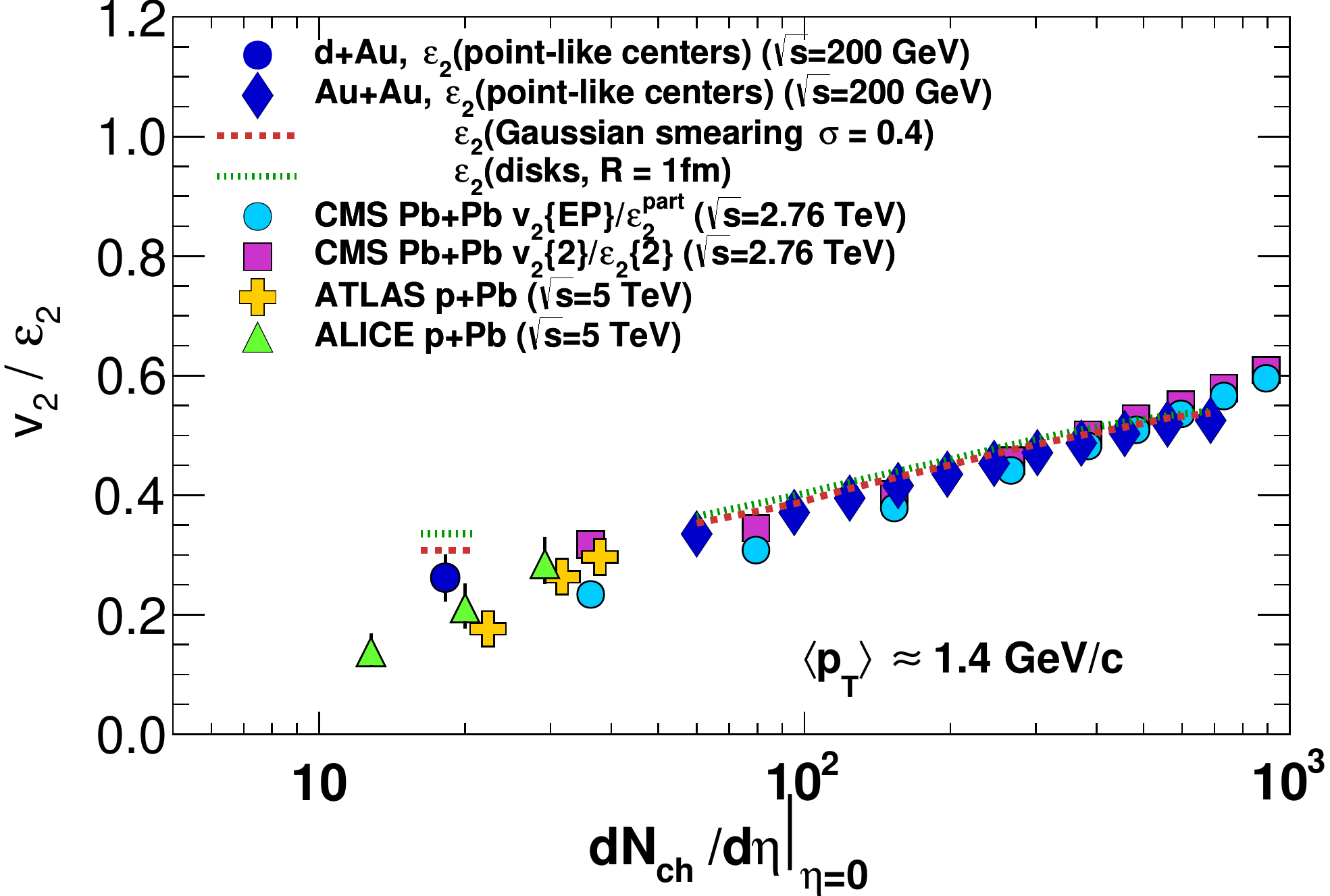}
\caption{$v_2/\varepsilon_2$ as a function of the midrapidity charged particle
multiplicity in d+Au, p+Pb, Au+Au and Pb+Pb collisions.  
The $\varepsilon_2$ values are calculated within a Glauber Monte Carlo.
Default $\varepsilon_2$ values are obtained representing the nucleons as point-like centers. Also
shown are results with the nucleons represented as Gaussians with $\sigma$~=~0.4~fm 
(red dashed line)
and solid disks with radius of 1~fm (green dotted line).
Figure is from Ref.~\cite{Adare:2013piz}.}
\label{fig:scaling}
\end{figure}

\begin{figure}
\centering
\includegraphics[width=0.7\textwidth]{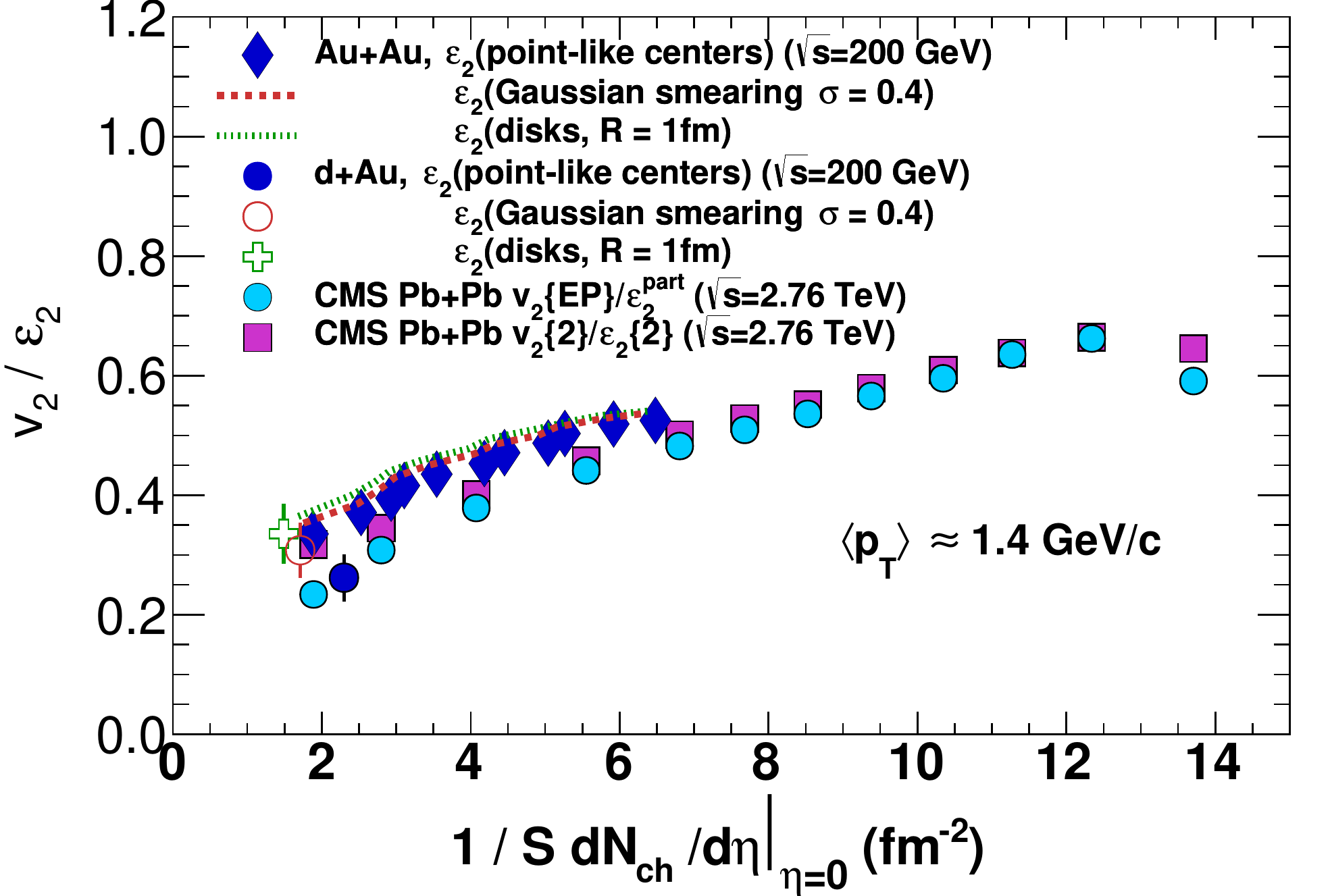}
\caption{$v_2/\varepsilon_2$ as a function of $\frac{1}{S} \frac{dN_{ch}}{d\eta}$
 in d+Au~\cite{Adare:2013piz}, 
Au+Au~\cite{Adler:2004zn,Lacey:2010fe} and Pb+Pb~\cite{Chatrchyan:2012ta} collisions.  
The $\varepsilon_2$ and $S$ values are calculated within a Glauber Monte Carlo.
Default values represent the nucleons as point-like centers. Also
shown for the Au+Au and d+Au points
 are results with the nucleons represented as Gaussians with $\sigma$~=~0.4~fm 
(red dashed line and red open circle)
and solid disks with radius of 1~fm (green dotted line and green open cross).}
\label{fig:areascaling}
\end{figure}

In order to further investigate the possible relationship between the geometry of the collision
system and the observed $v_2$ we compare the $v_2$ values (at $\pt~\approx$~1.4~GeV/c)
scaled by estimates of the initial state second order eccentricity, $\varepsilon_2$.
The $v_2/\varepsilon_2$ values are shown as a function of $dN_{ch}/d\eta$ at mid-rapidity
in Fig.~\ref{fig:scaling} for d+Au, p+Pb, Au+Au and Pb+Pb collisions at both RHIC and the LHC.  
The $\varepsilon_2$ values are from a Glauber Monte Carlo calculation;
$v_2/\varepsilon_2$ is consistent between central d+Au and midcentral p+Pb (systems
which have a similar $dN_{ch}/d\eta$) despite the factor of 25 difference in collision energy 
per nucleon pair.
The $\varepsilon_2$ value in central d+Au collisions is approximately 50\% larger than in
in midcentral p+Pb.
The $v_2/\varepsilon_2$ ratio
rises as a function $dN_{ch}/d\eta$ and follows approximately a common trend between
the four collisions systems.

There are uncertainties within the Glauber Monte Carlo calculation of the 
spatial eccentricity, $\varepsilon_2$, 
values.  We have investigated the uncertainty due to the modeling of the nucleons
within the calculation.  The default values take the nucleons as point-like centers and
we have also investigated treating the nucleons as disks with a radius of 1~fm and
as Gaussians with $\sigma$~=~0.4 in both d+Au and Au+Au collisions.  In d+Au collisions
these variations change $\varepsilon_2$ by approximately 30\% at maximum; in Au+Au the effect is much smaller.

In addition to the multiplicity scaling shown in Fig.~\ref{fig:scaling} it is also of interest
to investigate the scaling of the same $v_2 / \varepsilon_2$ data as a function of
the charged particle multiplicity divided by $S$, a measure of the overlap area
between the two nucleons which is defined as $S\equiv 4\pi\sqrt{\sigma_x^2\sigma_y^2 - \sigma_{xy}^2}$
where $\sigma_{xy}\equiv \langle xy \rangle - \langle x \rangle \langle y \rangle$.
This scaling has been found to approximately hold between RHIC and the LHC for $p_T$ integrated
$v_2/\varepsilon_2$ values~\cite{Chatrchyan:2012ta}.  In Fig.~\ref{fig:areascaling}
$v_2/ \varepsilon_2$ is plotted as a function of $\frac{1}{S}\frac{dN_{ch}}{d\eta}$
at $p_T\approx$~1.4~GeV/c.  We observe this scaling to hold in d+Au, Au+Au and Pb+Pb
collisions.  The same three assumptions
about the nucleon representation within the Glauber model are used here as well.  In this case,
the different nucleon representations affect both the values of $\varepsilon_2$ and $S$.

\section{Rapidity Separated Correlations}
\label{sec:rapsep}

In order to investigate whether the correlations seen at midrapidity are 
long-range in pseudorapidity and, if so, what the pseudorapidity dependence is, we have measured correlations between
mid-rapidity hadrons ($|\eta|<$0.35) and electromagnetic energy in the 
Au-going and d-going Muon Piston
Calorimeters (MPC)~\cite{Adare:2011sc}.  These calorimeters are positioned on either side of the interaction region
at 3.1~$<|\eta|<$~3.7 (3.9) in the Au-going (d-going) direction.  
With this large $\Delta\eta$ separation, no same-side jet correlations remain.

In d+Au collisions there is a large asymmetry in the number of particles produced
at large pseudorapidity between the $d$-going and Au-going directions.  In
the most central 20\% of the collisions approximately four times as many charged
particles are produced in the Au-going direction as in the $d$-going direction~\cite{Back:2004mr}.

\begin{figure}
\includegraphics[width=0.45\textwidth]{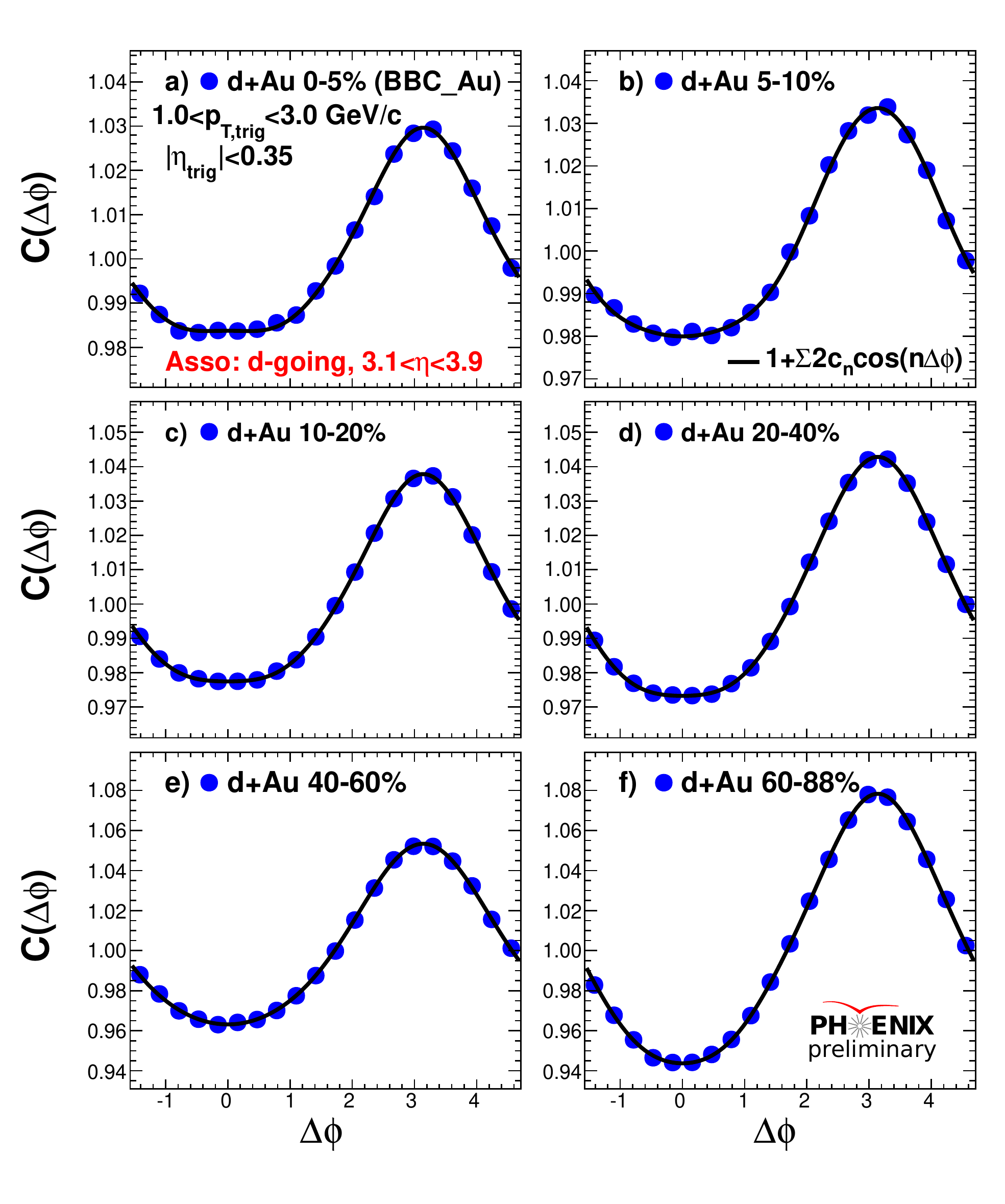}
\includegraphics[width=0.45\textwidth]{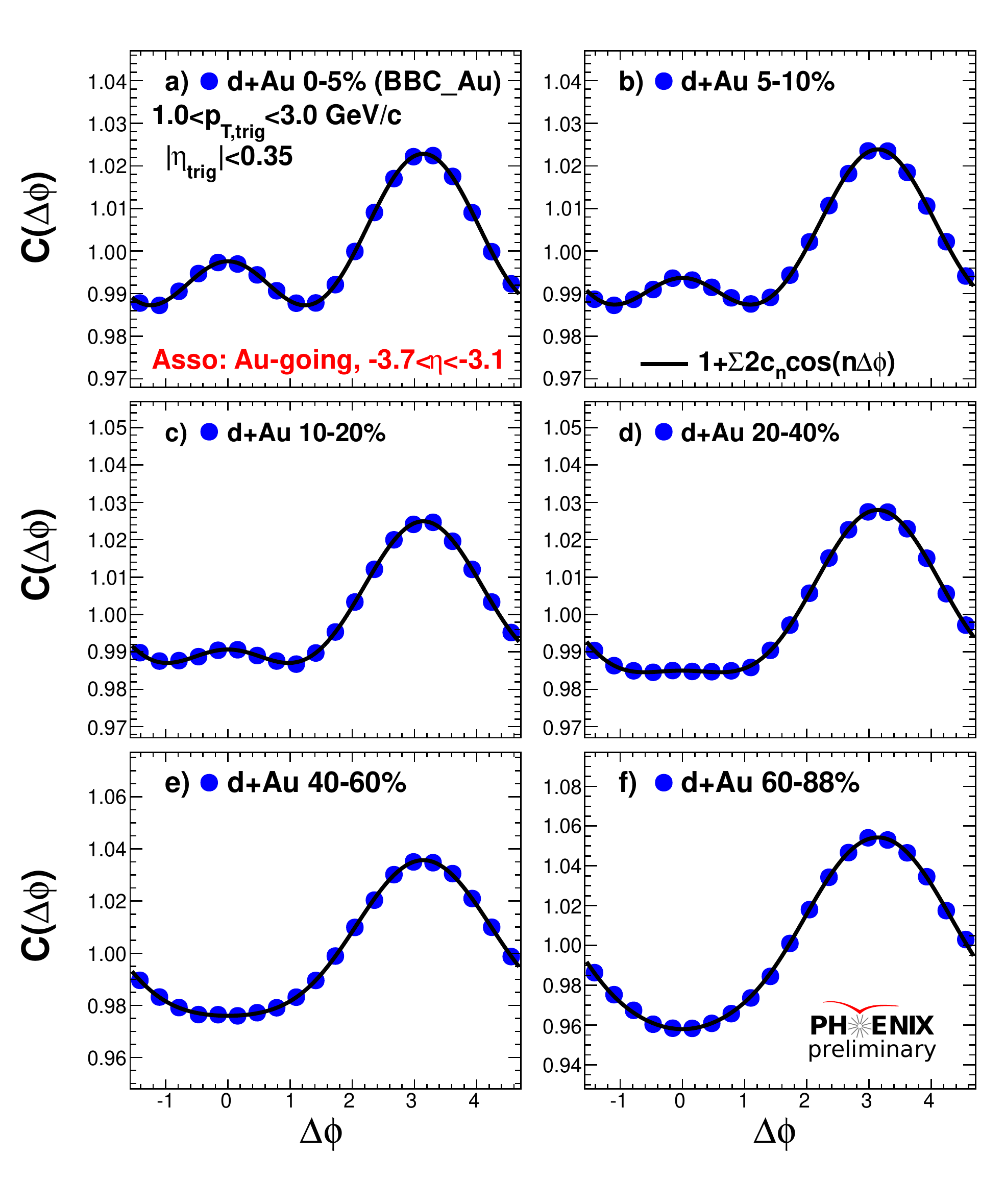}
\caption{Azimuthal correlations between mid-rapidity hadrons and $E_T$ as measured in the 
MPC in the d-going (left panels) and Au-going (right panels) direction. 
The centrality selections are as noted on the plots.}
\label{fig:mpc}
\end{figure}

The azimuthal correlations for both the d-going and Au-going MPC 
with midrapidity tracks are
shown in Fig.~\ref{fig:mpc} as a function of centrality.  In all cases the dominant
feature of the correlation functions is the peak at $\dphi$=$\pi$.  This peak
has contributions from jet correlations and momentum conservation.  

In the
d-going correlations no near side correlations are seen at any centrality.
In the Au-going correlations, peripheral correlations look similar to those
observed in the d-going direction.  However, in central collisions a 
positive correlation at $\Delta\phi=$~0 is observed for the 20\% most central events.
The magnitude of the correlation relative to the peak at $\dphi$=$\pi$
increases toward more central events. This provides evidence that the $v_2$
observed at midrapidity is from a long-range pseudorapidity correlation.

\section{Future Investigations}
\label{sec:future}

While many of the existing measurements in p+Pb and d+Au 
are suggestive of hydrodynamic behavior in d+Au collisions,
further investigations of this surprising effect are certainly in order. 
No significant $v_3$ in d+Au collisions was observed~\cite{Adare:2013piz}.  This is consistent
with the large $v_2$ driven by the elongated shape of the deuteron and, as discussed above, the $v_3$ 
values extracted from the d+Au data are much smaller than $v_2$ and consistent with zero.  
If the geometry
of the deuteron causes the observed $v_2$ values then it could be possible to induce a large $v_3$
by using a projectile nucleus with a large $\varepsilon_3$.  
PHENIX has proposed running $^3$He+Au, d+Au and p+Au collisions at RHIC in 2015
to directly investigate the relationship between $v_N$ and geometry.

\begin{figure}
\includegraphics[width=\textwidth]{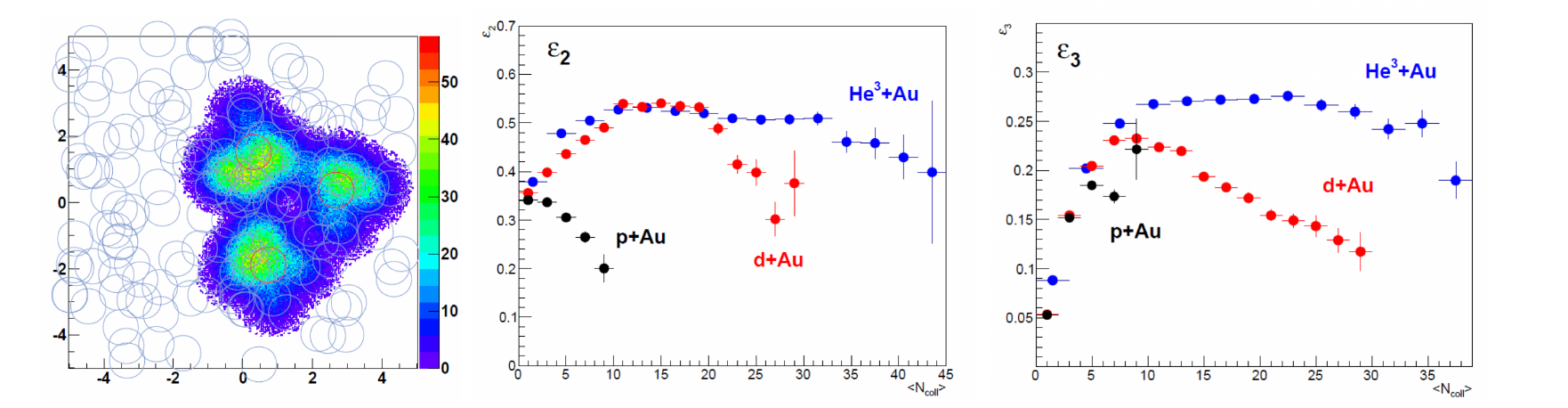}
\caption{(left)An event display of a $^3$He+Au collision as simulated in a Glauber Monte Carlo.
The nucleons are modeled as Gaussians with $\sigma$=0.4.  The outlines of the nucleons are shown as 
circles and the color axis shows the Glauber initial energy density. (middle) $\varepsilon_2$
as calculated in a Glauber Monte Carlo as a function of $N_{coll}$ for p+A (black), d+Au (red)
and $^3$He+Au (blue) collisions. (right) $\varepsilon_3$ as a function of $N_{coll}$.  Colors
are as in the middle panel.}
\label{fig:he3}
\end{figure}

Fig.~\ref{fig:he3} shows the results of Glauber Monte Carlo calculations of $^3$He+Au, d+Au and
p+Au collisions.  The $\varepsilon_2$ value decreases with an increasing number
of binary nucleon-nucleon collisions ($N_{coll}$) 
in p+Au collisions while for d+Au and $^3$He+Au collisions it increases
up to a maximal value of approximately 0.5.  The $\varepsilon_3$ in $^3$He+Au collisions reaches 0.25
at about 10 collisions and remains approximately constant for all more central collisions.  
In the 0-5\% central d+Au collisions studied here the mean $N_{coll}$ value is about 18, thus
the $\varepsilon_3$ value in central $^3$He+Au collisions will be approximately 40\% bigger than
in d+Au collisions.  

Additionally, in the 2015 run PHENIX will have additional tracking,
compared to the measurements presented here, provided by the silicon 
vertex detectors.  This will provides larger $\Delta\eta$ coverage around midrapidity.
Thus, it will be possible to determine whether the increased
$\varepsilon_3$ in $^3$He+Au collisions leads to a correspondingly large $v_3$ as a conclusive test
of the role of geometry in generating the $v_N$ observed in p+A and d+A collisions.
Recently, hydrodynamic calculations of $v_N$ in p+Au, d+Au and $^3$He+Au collisions have been 
done~\cite{Nagle:2013lja} and they provide support for pursing this program.

\section{Conclusions}

Over the past year, there has been much excitement around the novel effects observed in 
p+Pb and d+Au collisions.  We have presented PHENIX results on the d+Au $v_2$ at midrapidity
and shown new results using our MPC detector which provide the first evidence for long range 
($\Delta\eta \approx$~3.5)
correlations at small $\dphi$ in d+Au collisions.  We have also discussed future plans to 
constrain the role of geometry in small collision systems by varying the shape of the 
projectile nucleus in order to significantly vary the $\varepsilon_3$ value of the collision region.
The collision system variation possible at RHIC make feature studies in this direction
especially exciting.





\bibliographystyle{elsarticle-num.bst}
\bibliography{sickles_proceedings}

\begin{thebibliography}{10}
\expandafter\ifx\csname url\endcsname\relax
  \def\url#1{\texttt{#1}}\fi
\expandafter\ifx\csname urlprefix\endcsname\relax\def\urlprefix{URL }\fi
\expandafter\ifx\csname href\endcsname\relax
  \def\href#1#2{#2} \def\path#1{#1}\fi

\bibitem{Adare:2011tg}
A.~Adare, et~al., Phys.Rev.Lett. 107 (2011) 252301.
\newblock \href {http://arxiv.org/abs/1105.3928} {\path{arXiv:1105.3928}}.

\bibitem{Aamodt:2011by}
K.~Aamodt, et~al., Phys.Lett. B708 (2012) 249--264.
\newblock \href {http://arxiv.org/abs/1109.2501} {\path{arXiv:1109.2501}}.

\bibitem{ATLAS:2012at}
G.~Aad, et~al., Phys.Rev. C86 (2012) 014907.
\newblock \href {http://arxiv.org/abs/1203.3087} {\path{arXiv:1203.3087}}.

\bibitem{Abelev:2009af}
B.~Abelev, et~al., {Long range rapidity correlations and jet production in high
  energy nuclear collisions}, Phys.Rev. C80 (2009) 064912.
\newblock \href {http://arxiv.org/abs/0909.0191} {\path{arXiv:0909.0191}},
  \href {http://dx.doi.org/10.1103/PhysRevC.80.064912}
  {\path{doi:10.1103/PhysRevC.80.064912}}.

\bibitem{Khachatryan:2010gv}
V.~Khachatryan, et~al., JHEP 1009 (2010) 091.
\newblock \href {http://arxiv.org/abs/1009.4122} {\path{arXiv:1009.4122}}.

\bibitem{Abelev:2012ola}
B.~Abelev, et~al., Phys.Lett. B719 (2013) 29--41.
\newblock \href {http://arxiv.org/abs/1212.2001} {\path{arXiv:1212.2001}}.

\bibitem{Aad:2012gla}
G.~Aad, et~al., Phys.Rev.Lett. 110 (2013) 182302.
\newblock \href {http://arxiv.org/abs/1212.5198} {\path{arXiv:1212.5198}}.

\bibitem{Dusling:2012wy}
K.~Dusling, R.~Venugopalan, Phys.Rev. D87 (2013) 054014.
\newblock \href {http://arxiv.org/abs/1211.3701} {\path{arXiv:1211.3701}}.

\bibitem{Bozek:2011if}
P.~Bozek, Phys.Rev. C85 (2012) 014911.
\newblock \href {http://arxiv.org/abs/1112.0915} {\path{arXiv:1112.0915}}.

\bibitem{Bozek:privatecomm}
P.~Bozek, private communication.

\bibitem{Bzdak:2013zma}
A.~Bzdak, B.~Schenke, P.~Tribedy, R.~Venugopalan\href
  {http://arxiv.org/abs/1304.3403} {\path{arXiv:1304.3403}}.

\bibitem{Qin:2013bha}
G.-Y. Qin, B.~Muller\href {http://arxiv.org/abs/1306.3439}
  {\path{arXiv:1306.3439}}.

\bibitem{Adare:2013piz}
A.~Adare, et~al., Phys.Rev.Lett.\href {http://arxiv.org/abs/1303.1794}
  {\path{arXiv:1303.1794}}, \href
  {http://dx.doi.org/10.1103/PhysRevLett.111.212301}
  {\path{doi:10.1103/PhysRevLett.111.212301}}.

\bibitem{Adler:2004zn}
S.~Adler, et~al., Phys.Rev. C71 (2005) 034908.
\newblock \href {http://arxiv.org/abs/nucl-ex/0409015}
  {\path{arXiv:nucl-ex/0409015}}, \href
  {http://dx.doi.org/10.1103/PhysRevC.71.049901, 10.1103/PhysRevC.71.034908}
  {\path{doi:10.1103/PhysRevC.71.049901, 10.1103/PhysRevC.71.034908}}.

\bibitem{Lacey:2010fe}
R.~A. Lacey, A.~Taranenko, R.~Wei, N.~Ajitanand, J.~Alexander, et~al.,
  Phys.Rev. C82 (2010) 034910.
\newblock \href {http://arxiv.org/abs/1005.4979} {\path{arXiv:1005.4979}},
  \href {http://dx.doi.org/10.1103/PhysRevC.82.034910}
  {\path{doi:10.1103/PhysRevC.82.034910}}.

\bibitem{Chatrchyan:2012ta}
S.~Chatrchyan, et~al., Phys.Rev. C87 (2013) 014902.
\newblock \href {http://arxiv.org/abs/1204.1409} {\path{arXiv:1204.1409}},
  \href {http://dx.doi.org/10.1103/PhysRevC.87.014902}
  {\path{doi:10.1103/PhysRevC.87.014902}}.

\bibitem{Adare:2011sc}
A.~Adare, et~al., Phys.Rev.Lett. 107 (2011) 172301.
\newblock \href {http://arxiv.org/abs/1105.5112} {\path{arXiv:1105.5112}}.

\bibitem{Back:2004mr}
B.~Back, et~al., Phys.Rev. C72 (2005) 031901.
\newblock \href {http://arxiv.org/abs/nucl-ex/0409021}
  {\path{arXiv:nucl-ex/0409021}}, \href
  {http://dx.doi.org/10.1103/PhysRevC.72.031901}
  {\path{doi:10.1103/PhysRevC.72.031901}}.

\bibitem{Nagle:2013lja}
J.~Nagle, A.~Adare, S.~Beckman, T.~Koblesky, J.~O. Koop, et~al.\href
  {http://arxiv.org/abs/1312.4565} {\path{arXiv:1312.4565}}.

\end{thebibliography}







\end{document}